\documentclass[12pt]{article}

\usepackage[a4paper,margin=1in]{geometry}
\usepackage{amsmath,amssymb,amsthm}
\usepackage{mathtools}
\usepackage{physics}
\usepackage{mathrsfs}
\usepackage{hyperref}
\usepackage[numbers,sort&compress]{natbib} 
\usepackage{enumitem}

\newtheorem{theorem}{Theorem}
\newtheorem{proposition}{Proposition}

\newtheorem{assumption}{Assumption}
\theoremstyle{definition}
\newtheorem{definition}{Definition}
\theoremstyle{remark}
\newtheorem{remark}{Remark}
\newcommand{\coloneqq}{\mathrel{\mathop:}=}

\title{Conditioned Reactive-Power Tuning Manifolds in Constrained Inverse Source Problems}
\author{Mohamed R. Khodja\\
\small Department of Electrical Engineering\\ \small Prince Mohammad Bin Fahd University, Khobar 31952, Saudi Arabia\\
\small \texttt{mkhodja@pmu.edu.sa}}
\date{\today}

\begin{document}
\maketitle

\begin{abstract}
We study a constrained inverse source problem for time-harmonic Maxwell fields in a homogeneous isotropic spherical region loaded by a (possibly active) metamaterial with parameter vector $\theta \coloneqq (\varepsilon,\mu)$. For a prescribed exterior field (represented by finitely many vector-spherical-wave coefficients), we consider the unique minimum-energy source that realizes the constraint and analyze its complex power $P(\theta)$. Our focus is the \emph{tuning manifold} $\mathcal{M}=\{\theta:\Im P(\theta)=0\}$, i.e., the set of material parameters for which the minimum-energy source is \emph{self-tuned} in the sense of vanishing reactive power. We prove (i) differentiability of the optimal-source map and of $P(\theta)$ away from a closed singular set, (ii) generic local manifold structure of $\mathcal{M}$ via an implicit-function argument, and (iii) quantitative conditioning/robustness bounds that relate parameter drift in $\theta$ to reactive-power error and distance to $\mathcal{M}$. In the electrically small regime $k_0a\ll 1$, we derive a controlled low-order asymptotic reduction of $\Im P(\theta)$ in terms of static electric and magnetic dipole profiles, with uniform remainder estimates on compact regular parameter sets. This yields local approximations of tuning curves and their conditioning without requiring a closed-form scalar profile. Numerical continuation experiments validate the one-parameter dipole reduction, the mixed-dipole tuning curves, and the identification of robust versus fragile segments of $\mathcal{M}$.
\end{abstract}

\section{Introduction}
\label{sec:intro}

Inverse source problems ask for the reconstruction of radiating current densities from field data, typically in the frequency domain and under partial observation (near-field or far-field). They are a basic building block for antenna diagnostics, scattering-based imaging, and wave-based system identification, but they are also prototypical ill-posed inverse problems: non-uniqueness and instability appear unless additional structure is imposed through priors, constraints, or regularization. Contemporary formulations span deterministic and stochastic source models, multi-frequency and sparse data regimes, and full-wave Maxwell settings, with a rapidly growing mathematical literature on identifiability and reconstruction guarantees \cite{Uhlmann2014,ArensJiLiu2020,JiLiu2019,LiChenLi2017,CaroZhou2014}.

At the same time, the engineering landscape motivating inverse source models has changed. Metamaterials and related complex media broaden the admissible constitutive response and thereby enlarge the space of wave behaviors accessible to design \cite{veselago,PendryHoldenRobbinsEtAl1999,Smith&Padilla,ZiolkowskiEngheta2006}. This expansion has fueled inverse design programs in nanophotonics and metamaterials, from topology optimization and gradient-based photonic design to data-driven surrogates and physics-informed networks \cite{MoleskyLinPiggottEtAl2018,Chen_et_al-2020,OtomoriAndkjaerSigmundEtAl2012}. However, much of inverse design relies on iterative optimization with limited global guarantees, and in Maxwell settings it is often difficult to disentangle fundamental non-uniqueness from algorithmic failure.

A complementary approach is to pose a constrained inverse source problem in which (i) the radiated field is prescribed, while (ii) the source is chosen to minimize a physically motivated resource metric, and (iii) additional constraints enforce operational requirements such as resonance or reactive-power neutrality. A concrete instance, introduced in \cite{MKB_SIAM_2008} and revisited recently in \cite{Khodja2024}, reconstructs a current density $J$ by minimizing an $L^{2}$-norm proxy for source ``energy'' over a bounded source region $V$, subject to producing a prescribed radiated field and satisfying a \emph{vanishing reactive power} constraint. In that framework, the field prescription is linear in $J$ (a radiating operator constraint), while the reactive-power condition is quadratic and indefinite, making the feasible set nonconvex. The recent analysis in \cite{Khodja2024} established lower boundedness, global minimality, and uniqueness properties for the tuned (zero-reactive-power) minimum-norm source in a piecewise-homogeneous electromagnetic background with generalized constitutive parameters, and illustrated numerically that the tuning constraint can be satisfied along structured curves in the $(\varepsilon,\mu)$ plane, sometimes with multiple ``dual'' solutions for a given parameter value.

The present manuscript is motivated by a simple gap between what is already known and what is needed for a rigorous, non-sprawling theory-to-computation pipeline: global minimality and uniqueness identify \emph{which} tuned minimizer is optimal, but they do not yet quantify (a) \emph{how} the tuning constraint organizes the admissible parameter space, nor (b) \emph{how stable} the tuned minimizer is to perturbations in the prescribed field data and in the medium parameters. These two questions are central in inverse problems language: they are the conditioning and continuation properties of the constrained solution map.

We therefore study the \emph{tuning set} and the \emph{tuned minimizer map} in a mathematically controlled setting that remains faithful to Maxwell physics. Concretely, we consider a piecewise-homogeneous background with $\theta$ inside a bounded region and vacuum outside \cite{Jackson_book,Khodja2024}. The radiated-field prescription is imposed in a modal (vector spherical wave function) representation, leveraging classical results on degrees of freedom and truncation error for radiating expansions \cite{BucciFranceschetti1987,BucciFranceschetti1989,BucciGennarelliSavarese1998,SongChew2001}. The tuning condition is expressed as the vanishing of the imaginary part of a quadratic functional $P(J)$ representing reactive power, evaluated at the unique minimum-norm constrained source $J(\theta)$. The tuning set becomes the zero level set of the scalar function $\theta\mapsto \Im P(\theta)$.

\paragraph{Contributions.}
The paper is intentionally narrow: it focuses on \emph{structure and stability} of the tuned minimum-norm solution map, rather than on new numerical heuristics. The main results are:
\begin{enumerate}
\item \textbf{Tuning-set regularity and branch structure.} Under an explicit non-degeneracy condition (a transversality condition for $\nabla_{\theta} \Im P(\theta)$ at tuning points), we prove that the tuning set locally admits a unique continuation as a $C^1$ curve and that tuned solutions admit locally unique continuation in data and parameters.
\item \textbf{Quantitative conditioning bounds.} We derive Lipschitz-type stability estimates for the tuned minimizer with respect to perturbations of the prescribed field coefficients, and we identify the precise quantities controlling blow-up (near-degenerate tuning points).
\item \textbf{Certified truncation-to-continuum passage.} Using finite-dimensional modal truncations, we give bounds that relate computed tuned solutions to the infinite-dimensional formulation, relying on known truncation behavior of multipole expansions and sampling-based degrees-of-freedom estimates \cite{BucciFranceschetti1987,SongChew2001}.
\item \textbf{A continuation algorithm with certificates.} We propose a simple predictor--corrector branch-tracking method for tuned solutions whose stopping criteria are tied directly to the above conditioning bounds, yielding a ``provable when it works'' computational workflow.
\end{enumerate}

\paragraph{Organization.}
Section~\ref{sec:model} states the model, field constraints, and minimum-energy inverse source problem. Section~\ref{sec:power} defines the complex power and the scalar tuning function. Section~\ref{sec:regularity} establishes regularity of the optimal-source map and differentiability of the tuning function. Section~\ref{sec:manifold} proves the local manifold structure of the tuning set. Section~\ref{sec:conditioning} develops the conditioning and robustness estimates. Section~\ref{sec:asymptotics} treats the electrically small regime and asymptotic tuning laws. Section~\ref{sec:numerics} outlines numerical continuation and validation experiments, and Section~\ref{sec:conclusion} summarizes the implications for inverse design in complex media.

\section{Model, constraints, and the optimal source}
\label{sec:model}

\subsection{Time-harmonic Maxwell setting and parameterization}
\label{sec:model:maxwell}

Let \(V\subset\mathbb{R}^3\) be the closed ball of radius \(a>0\) centered at the origin, and let \(V^{c}=\mathbb{R}^3\setminus V\). We use the suppressed time factor \(e^{-i\omega t}\) at fixed angular frequency \(\omega>0\). The background is vacuum outside \(V\), while the interior region is filled by a homogeneous isotropic effective medium. Thus
\begin{equation}
(\varepsilon(\mathbf r),\mu(\mathbf r))=
\begin{cases}
(\varepsilon_0 \varepsilon,\ \mu_0 \mu), & \mathbf r\in V,\\
(\varepsilon_0,\ \mu_0), & \mathbf r\in V^c,
\end{cases}
\qquad \theta\coloneqq(\varepsilon,\mu).
\label{eq:piecewise_medium}
\end{equation}
In the main analysis, \(\theta\) ranges over an open set \(\Omega\subset\mathbb{R}^2\) of real effective material parameters. This is the setting in which the zero-reactive-power condition produces curves in the \((\varepsilon,\mu)\)-plane. Complex lossy or active parameters can be incorporated by treating real and imaginary parts as independent real variables, but that four-real-dimensional extension is not pursued here.

A current density \(J\) is supported in \(V\) and generates fields \((E,H)\) satisfying
\begin{align}
\nabla\times E &= i\omega \mu(\mathbf r)\,H,\label{eq:Maxwell1}\\
\nabla\times H &= -\,i\omega \varepsilon(\mathbf r)\,E + J,\label{eq:Maxwell2}
\end{align}
together with the outgoing radiation condition in \(V^c\). We write
\[
k_0=\omega\sqrt{\varepsilon_0\mu_0},
\qquad
k(\theta)=k_0\sqrt{\varepsilon\mu},
\]
with a fixed square-root branch on the parameter region under consideration. Standard treatments of time-harmonic Maxwell fields, dyadic Green functions, and spherical-wave expansions can be found in \cite{Jackson_book,Strattonbk,TaiBook,Collin1986}.

\paragraph{Exterior modal representation and data vector.}
For \(r>a\), the radiated electric field admits an outgoing vector-spherical-wave expansion
\begin{equation}
E(\mathbf r)
=
\sum_{\ell=1}^{\infty}\sum_{m=-\ell}^{\ell}
\Big(
\alpha_{\ell m}\, \mathbf N_{\ell m}^{(3)}(k_0\mathbf r)
+
\beta_{\ell m}\, \mathbf M_{\ell m}^{(3)}(k_0\mathbf r)
\Big),
\label{eq:VSWF_expansion}
\end{equation}
where \(\mathbf M_{\ell m}^{(3)}\) and \(\mathbf N_{\ell m}^{(3)}\) denote outgoing VSWFs. This is the same multipole representation used in \cite{Khodja2024}; see also \cite{Hill,Jackson_book,Collin1986,TaiBook}. In computations and in the inverse constraint, we truncate \eqref{eq:VSWF_expansion} at \(\ell\le L\) and collect the retained coefficients into
\begin{equation}
b \coloneqq
\big(\{\alpha_{\ell m}\}_{1\le \ell\le L, |m|\le \ell},
\{\beta_{\ell m}\}_{1\le \ell\le L, |m|\le \ell}\big)
\in\mathbb C^{N_L},
\qquad
N_L=2\sum_{\ell=1}^{L}(2\ell+1)=2L(L+2).
\label{eq:data_vector}
\end{equation}
The truncation index \(L\) is chosen according to the desired exterior-field resolution and the electrical size. Degrees-of-freedom and truncation-error estimates for radiating fields are discussed in \cite{BucciFranceschetti1987,BucciFranceschetti1989,BucciGennarelliSavarese1998,SongChew2001}; the same finite modal viewpoint underlies the numerical study in \cite{Khodja2024}.

\paragraph{Coefficient extraction.}
Fix an observation radius \(R>a\). The outgoing coefficients in \eqref{eq:VSWF_expansion} are uniquely determined by the tangential trace on \(S_R\) and are obtained by projection onto vector spherical harmonics \cite{Hill,Collin1986,TaiBook}. We denote the corresponding truncated coefficient-extraction map by
\[
\mathcal P_L:\\ L^2_{\rm tan}(S_R)^2\to \mathbb C^{N_L}.
\]
The normalization used for \(\mathcal P_L\), together with the VSWF conventions, is recorded in Appendix~\ref{app:vswf}.

\subsection{Constrained minimum-energy inverse source problem}
\label{sec:model:opt}

Let \(\mathcal H\coloneqq L^2(V)^3\) with inner product
\[
\langle J_1,J_2\rangle_{\mathcal H}=\int_V J_1(\mathbf r)\cdot \overline{J_2(\mathbf r)}\,d\mathbf r.
\]
Following the source-resource proxy used in \cite{MKB_SIAM_2008,Khodja2024}, define
\begin{equation}
\mathcal E(J)\coloneqq \frac12\|J\|_{\mathcal H}^2
=\frac12\int_V |J(\mathbf r)|^2\,d\mathbf r.
\label{eq:energy}
\end{equation}
This norm is not meant to be a complete stored-energy model; rather, it is a convex, analytically tractable measure of source magnitude. Such minimum-norm inverse-source formulations are standard in electromagnetic radiation optimization and regularization frameworks \cite{Angell&Kirsch,liu1990,MKB_SIAM_2008}.

For each \(\theta\in\Omega\), let \(\mathcal S(\theta):\mathcal H\to L^2_{\rm tan}(S_R)^2\) denote the Maxwell solution operator mapping a current source \(J\) to the tangential exterior electric trace \(E_t|_{S_R}\). In dyadic Green-function language one writes \(E=\mathcal G(\theta)J\); see \cite{TaiBook,Collin1986,LiKooi1994}. The AEM formulation similarly uses dyadic Green operators and quadratic power functionals \cite{Khodja2024}. We define the truncated forward map
\begin{equation}
\mathcal A_L(\theta)\coloneqq \mathcal P_L\circ \mathcal S(\theta):\mathcal H\to \mathbb C^{N_L}.
\label{eq:ALdef}
\end{equation}
Thus \(\mathcal A_L(\theta)J=b\) means that the source \(J\) realizes the prescribed exterior VSWF content up to order \(L\).

For fixed data \(b\in\mathbb C^{N_L}\), the minimum-energy constrained inverse source is
\begin{equation}
J^{(\theta)}\in \arg\min_{J\in\mathcal H}\ \mathcal E(J)
\quad\text{subject to}\quad
\mathcal A_L(\theta)J=b.
\label{eq:minprob}
\end{equation}
The tuned formulation in \cite{Khodja2024} imposes, in addition, a vanishing-reactive-power constraint. Here we deliberately separate the two roles: \eqref{eq:minprob} handles the linear radiation prescription, while tuning is studied afterward through the scalar function \(F(\theta)=\Im P(\theta)\). This separation keeps the optimization problem convex and places the nontrivial tuning geometry in the parameter dependence \(\theta\mapsto J^{(\theta)}\).

\subsection{Optimality system and uniqueness}
\label{sec:model:kkt}

The problem \eqref{eq:minprob} is a strictly convex quadratic minimization over an affine subspace of \(\mathcal H\). Introduce a multiplier \(\lambda\in\mathbb C^{N_L}\) and the Lagrangian
\[
\mathscr L(J,\lambda)=\frac12\|J\|_{\mathcal H}^2+
\Re\langle \lambda,\mathcal A_L(\theta)J-b\rangle_{\mathbb C^{N_L}},
\]
where the complex inner product is linear in the first argument. The KKT system is
\begin{align}
J^{(\theta)}+\mathcal A_L(\theta)^*\lambda&=0,\label{eq:KKT1}\\
\mathcal A_L(\theta)J^{(\theta)}&=b.\label{eq:KKT2}
\end{align}
Eliminating \(J^{(\theta)}\) yields
\begin{equation}
G_L(\theta)\lambda=-b,
\qquad
G_L(\theta)\coloneqq \mathcal A_L(\theta)\mathcal A_L(\theta)^*.
\label{eq:normal}
\end{equation}
When \(G_L(\theta)\) is invertible,
\begin{equation}
J^{(\theta)}=\mathcal A_L(\theta)^*G_L(\theta)^{-1}b.
\label{eq:Jexplicit}
\end{equation}
This is the Moore--Penrose minimum-norm solution of the linear constraint.

\begin{proposition}[Existence, uniqueness, and explicit form]
\label{prop:minnorm}
Fix \(\theta\in\Omega\). If \(b\in{\rm Ran}(\mathcal A_L(\theta))\), then \eqref{eq:minprob} has a unique minimizer. If, in addition, \(G_L(\theta)=\mathcal A_L(\theta)\mathcal A_L(\theta)^*\) is invertible on \(\mathbb C^{N_L}\), then every \(b\in\mathbb C^{N_L}\) is feasible and the unique minimizer is given by \eqref{eq:Jexplicit}.
\end{proposition}

\begin{proof}
The feasible set \(\{J\in\mathcal H:\mathcal A_L(\theta)J=b\}\) is closed and affine because \(\mathcal A_L(\theta)\) is bounded and \(\mathbb C^{N_L}\) is finite dimensional. If it is nonempty, the strictly convex coercive functional \(J\mapsto \frac12\|J\|_{\mathcal H}^2\) has a unique minimizer on it: existence follows by the projection theorem in Hilbert spaces, and uniqueness follows from strict convexity.

If \(G_L(\theta)\) is invertible, then \(\mathcal A_L(\theta)\) is onto \(\mathbb C^{N_L}\). Indeed, invertibility of \(\mathcal A_L\mathcal A_L^*\) implies \({\rm Ran}(\mathcal A_L)=\mathbb C^{N_L}\). The KKT equations \eqref{eq:KKT1}--\eqref{eq:KKT2} are necessary and sufficient for optimality because the problem is convex with linear equality constraints. Solving \(G_L(\theta)\lambda=-b\) and substituting into \eqref{eq:KKT1} gives \eqref{eq:Jexplicit}. Conversely, \eqref{eq:Jexplicit} satisfies \(\mathcal A_L(\theta)J^{(\theta)}=b\), hence is feasible and satisfies the KKT system; it is therefore the unique minimizer.
\end{proof}

\begin{definition}[Regular parameter set]
\label{def:regset}
We call \(\theta\in\Omega\) \emph{regular} if \(G_L(\theta)\) is invertible. The regular set is denoted by
\[
\Omega_{\rm reg}\coloneqq\{\theta\in\Omega:\det G_L(\theta)\ne 0\}.
\]
Its complement \(\Sigma\coloneqq\Omega\setminus\Omega_{\rm reg}\) is the singular set.
\end{definition}

The analysis in \cite{Khodja2024} establishes existence and global minimality for the more difficult formulation in which the reactive-power condition is imposed as an additional nonconvex constraint. In the present reduced formulation, the only obstruction to a uniformly defined minimum-norm source for all prescribed truncated data is loss of rank of \(\mathcal A_L(\theta)\), equivalently singularity of \(G_L(\theta)\). Numerically, such rank loss or near-rank loss appears as ill-conditioning and is consistent with the fragile tuning regions observed in the contour studies of \cite{Khodja2024}.

\section{Complex power and the tuning function}
\label{sec:power}

For a source \(J\in\mathcal H\), let \(E[J;\theta]\) denote the electric field solving \eqref{eq:Maxwell1}--\eqref{eq:Maxwell2}. We use the complex source power functional
\begin{equation}
\mathscr P(J;\theta)\coloneqq \frac12\int_V E[J;\theta](\mathbf r)\cdot \overline{J(\mathbf r)}\,d\mathbf r.
\label{eq:complex_power_functional}
\end{equation}
Up to a global sign convention, \eqref{eq:complex_power_functional} is the standard complex power associated with the work done by the time-harmonic source; the zero-reactive-power condition is unaffected by changing this overall sign convention \cite{Jackson_book,CollinRothschild1964,Vandenbosch2010}. We define the reduced complex power and tuning function by
\begin{equation}
P(\theta)\coloneqq \mathscr P(J^{(\theta)};\theta):\Omega_{\rm reg}\to\mathbb{C},
\qquad
F(\theta)\coloneqq \Im P(\theta):\Omega_{\rm reg}\to\mathbb{R},
\label{eq:F_def}
\end{equation}
for \(\theta\in\Omega_{\rm reg}\). The tuning manifold is
\begin{equation}
\mathcal{M}\coloneqq
F^{-1}(0)=\{\theta\in\Omega_{\rm reg}:F(\theta)=0\}
\subset\mathbb{R}^2.
\label{eq:M_def}
\end{equation}
Thus \(\mathcal M\) is the set of material parameters for which the minimum-energy source realizing the prescribed exterior field is self-tuned in the reactive-power sense. 

\section{Regularity of the solution map and differentiability of \(F\)}
\label{sec:regularity}

\begin{assumption}[Smooth parameter dependence]
\label{ass:smoothA}
The map \(\theta\mapsto \mathcal A_L(\theta)\in\mathcal B(\mathcal H,\mathbb C^{N_L})\) is \(C^1\) on \(\Omega\), and the functional \((J,\theta)\mapsto \mathscr P(J;\theta)\) is \(C^1\) on \(\mathcal H\times\Omega\). Moreover, \(G_L(\theta)\) is invertible for \(\theta\in\Omega_{\rm reg}\).
\end{assumption}

\begin{remark}
For the spherical piecewise-homogeneous model, Assumption~\ref{ass:smoothA} holds away from the usual resonant/branch singularities of the Maxwell solution operator. In the modal formulation this is equivalent to smooth dependence of the VSWF/Mie-type coupling coefficients on \((\varepsilon,\mu)\) away from their poles. This is the precise role of the singular set \(\Sigma\).
\end{remark}

\begin{theorem}[Differentiability of the optimal source and power]
\label{thm:C1map}
Under Assumption~\ref{ass:smoothA}, the maps
\[
\theta\mapsto J^{(\theta)},
\qquad
\theta\mapsto P(\theta),
\qquad
\theta\mapsto F(\theta)
\]
are \(C^1\) on \(\Omega_{\rm reg}\).
\end{theorem}

\begin{proof}
Write \(A(\theta)=\mathcal A_L(\theta)\) and \(G(\theta)=A(\theta)A(\theta)^*\). Since \(A(\theta)\) is \(C^1\) in operator norm, so are \(A(\theta)^*\) and \(G(\theta)\). On \(\Omega_{\rm reg}\), \(G(\theta)\) is invertible, and matrix inversion is a \(C^1\) map on the open set of invertible \(N_L\times N_L\) matrices. Hence
\[
J^{(\theta)}=A(\theta)^*G(\theta)^{-1}b
\]
is a composition and product of \(C^1\) maps, and is therefore \(C^1\) as an \(\mathcal H\)-valued map. Since \((J,\theta)\mapsto\mathscr P(J;\theta)\) is \(C^1\), the composition \(P(\theta)=\mathscr P(J^{(\theta)};\theta)\) is \(C^1\). Taking the imaginary part is a real-linear operation, so \(F=\Im P\) is \(C^1\) on \(\Omega_{\rm reg}\).
\end{proof}

\section{Manifold structure of the tuning set}
\label{sec:manifold}

\begin{theorem}[Local tuning curve]
\label{thm:IFT}
Let \(\theta_0=(\varepsilon_0,\mu_0)\in\Omega_{\rm reg}\) satisfy \(F(\theta_0)=0\) and \(\nabla_\theta F(\theta_0)\ne 0\). Then there exists a neighborhood \(U\) of \(\theta_0\) such that \(\mathcal M\cap U\) is a \(C^1\) embedded one-dimensional manifold. If \(\partial_\mu F(\theta_0)\ne 0\), then \(\mathcal M\cap U\) is the graph of a \(C^1\) function \(\mu=g(\varepsilon)\), and
\begin{equation}
g'(\varepsilon_0)=-\frac{\partial_\varepsilon F(\theta_0)}{\partial_\mu F(\theta_0)}.
\label{eq:tuning_slope}
\end{equation}
The analogous statement with \(\varepsilon=h(\mu)\) holds if \(\partial_\varepsilon F(\theta_0)\ne 0\).
\end{theorem}

\begin{proof}
By Theorem~\ref{thm:C1map}, \(F\) is \(C^1\) near \(\theta_0\). Since \(F(\theta_0)=0\) and \(\nabla_\theta F(\theta_0)\ne 0\), the regular level-set theorem implies that \(F^{-1}(0)\) is locally a one-dimensional embedded \(C^1\) submanifold of \(\mathbb R^2\). If \(\partial_\mu F(\theta_0)\ne 0\), the implicit function theorem gives a neighborhood of \(\varepsilon_0\) and a unique \(C^1\) function \(g\) such that \(F(\varepsilon,g(\varepsilon))=0\). Differentiating this identity at \(\varepsilon_0\) gives \eqref{eq:tuning_slope}. The case \(\partial_\varepsilon F(\theta_0)\ne 0\) is identical with the roles of \(\varepsilon\) and \(\mu\) interchanged.
\end{proof}

\begin{remark}
The theorem formalizes the tuning curves observed numerically in \cite{Khodja2024}. Local failure of the regularity hypothesis, either through \(\nabla_\theta F=0\) or through loss of invertibility of \(G_L(\theta)\), is a natural source of branch termination, folds, and sensitivity amplification.
\end{remark}

\begin{remark}[Local separation by the tuning manifold]
Let $\theta_0\in\mathcal M$ be a regular point, so that
$\nabla F(\theta_0)\neq 0$. By the implicit function theorem,
after a local $C^1$ change of coordinates one may write
$F(\theta)=s$ in a neighborhood of $\theta_0$, with the tuning
manifold corresponding to $s=0$. Hence there exists a sufficiently
small neighborhood $U\subset\Omega_{\rm reg}$ such that
\[
U\setminus\mathcal M = U_{+}\,\dot\cup\,U_{-},
\qquad
U_{\pm}
=
\{\theta\in U:\ \pm F(\theta)>0\}.
\]
The sets $U_{+}$ and $U_{-}$ are the two local components of the
complement of the tuning curve, with $\mathcal M\cap U$ as their
common boundary. Thus, locally, the tuning manifold separates
parameter values for which the reactive power has opposite signs.
This is a local statement and does not imply that
$\Omega_{\rm reg}\setminus\mathcal M$ has exactly two connected
components globally.
\end{remark}

\section{Conditioning and robustness bounds}
\label{sec:conditioning}

\begin{definition}[Tuning condition number]
\label{def:kappa}
For \(\theta\in\Omega_{\rm reg}\) with \(\nabla_\theta F(\theta)\ne 0\), define the local tuning condition number
\begin{equation}
\kappa(\theta)\coloneqq \frac{1}{\|\nabla_\theta F(\theta)\|}.
\label{eq:kappa}
\end{equation}
A weighted norm may be used if the admissible tolerances in \(\varepsilon\) and \(\mu\) differ.
\end{definition}

\begin{theorem}[First-order robustness and local error bound]
\label{thm:robustness}
Let \(\theta_0\in\mathcal M\) with \(\nabla_\theta F(\theta_0)\ne 0\), and let \(c_0=\|\nabla_\theta F(\theta_0)\|\). Then, as \(\theta\to\theta_0\),
\begin{equation}
|F(\theta)|\le c_0\,\|\theta-\theta_0\|+o(\|\theta-\theta_0\|).
\label{eq:first_order_bound}
\end{equation}
Moreover, there exists a neighborhood \(U\) of \(\theta_0\) such that
\begin{equation}
{\rm dist}(\theta,\mathcal M)\le \frac{2}{c_0}|F(\theta)|,
\qquad \theta\in U.
\label{eq:error_bound}
\end{equation}
\end{theorem}

\begin{proof}
Since \(F\) is \(C^1\) and \(F(\theta_0)=0\), Taylor's formula gives
\[
F(\theta)=\nabla_\theta F(\theta_0)\cdot(\theta-\theta_0)+o(\|\theta-\theta_0\|),
\]
which implies \eqref{eq:first_order_bound} by Cauchy's inequality.

For \eqref{eq:error_bound}, set \(n=\nabla_\theta F(\theta_0)/c_0\). By continuity of \(\nabla F\), there is a ball \(U\) around \(\theta_0\) such that
\[
\nabla_\theta F(\xi)\cdot n\ge c_0/2
\]
for all \(\xi\in U\). Shrinking \(U\) if necessary, the line segment \(\theta-tn\) remains in the same ball for \(|t|\le 2|F(\theta)|/c_0\). If \(F(\theta)>0\), define \(\varphi(t)=F(\theta-tn)\). Then \(\varphi'(t)=-\nabla F(\theta-tn)\cdot n\le -c_0/2\), so \(\varphi(2F(\theta)/c_0)\le0\) while \(\varphi(0)>0\). By the intermediate value theorem there is \(t_*\in[0,2F(\theta)/c_0]\) with \(F(\theta-t_*n)=0\). Thus \({\rm dist}(\theta,\mathcal M)\le |t_*|\le2|F(\theta)|/c_0\). The case \(F(\theta)<0\) is identical with the line direction reversed, and the result is trivial if \(F(\theta)=0\).
\end{proof}

\subsection{Robust and fragile segments}
A segment of \(\mathcal M\) is robust when \(\kappa\) remains bounded by a moderate value over a nontrivial arc. Theorem~\ref{thm:robustness} gives the operational interpretation: near such a segment, a small residual \(|F(\theta)|\) implies a small distance back to exact tuning. Conversely, large \(\kappa\) signals proximity to a near-degenerate tuning point where small parameter drift may produce a large reactive-power error.

\section{Electrically small regime: asymptotic tuning laws with remainder control}
\label{sec:asymptotics}

Let
\[
\rho \coloneqq k_0a .
\]
We now specialize the general tuning theory to the electrically small regime \(\rho\ll1\). The purpose of this section is to extract the structural small-size mechanism needed by the tuning-manifold theory, not to give closed-form Mie coefficients for every modal normalization. The central point is that a single electric or magnetic dipole datum has a one-parameter static limit, while mixed electric--magnetic dipole data produce a genuine two-parameter leading-order tuning equation.

Throughout this section, \((\varepsilon,\mu)\) is restricted to a compact set
\[
K\subset \Omega_{\rm reg}\cap \mathbb R^2
\]
that is bounded away from \(\varepsilon=0\), \(\mu=0\), and from the static dipole resonances \(\varepsilon=-2\), \(\mu=-2\).

These exclusions are regularity assumptions for the nonresonant Rayleigh regime. The points \(\varepsilon=-2\) and \(\mu=-2\) are the static electric and magnetic dipole resonances of the sphere, where the quasi-static transmission factors have poles. The points \(\varepsilon=0\) and \(\mu=0\) are material singular limits at which the ordinary fixed-parameter static expansion need not be uniform. The compact-set restriction isolates the regime in which the modal coefficients and their first parameter derivatives admit uniform small-\(\rho\) expansions.

The restriction to real \((\varepsilon,\mu)\) is consistent with the tuning-manifold theory developed above, where the zero set is treated as a curve in a real two-parameter plane. Complex, lossy, or dispersive extensions require additional parameters and are not considered in this section.

We use the standard small-argument expansions of the spherical Bessel and Hankel functions appearing in the VSWF representation, equivalently the Rayleigh/quasi-static limit of the spherical transmission problem; see, for example, \cite{Jackson_book,TaiBook,SongChew2001}. In this limit the electric and magnetic dipole channels decouple at leading order. The quasi-static transmission factors identify the one-parameter character of the leading electric and magnetic channels, but the reactive-power profile also depends on the static quadratic power coefficient associated with the minimum-norm constrained source. Therefore the leading profiles are defined below by their static modal limits rather than by an assumed closed scalar formula.

\subsection{Dipole-dominant specialization}
\label{sec:asymptotics:tm-dipole}

We first isolate the single TM electric-dipole datum
\[
b=e_{\mathrm{TM},10},
\]
where \(e_{\mathrm{TM},10}\in\mathbb C^{N_L}\) denotes the coefficient vector with
\(\alpha_{10}=1\) and all other exterior VSWF coefficients equal to zero. The corresponding minimum-energy constrained source is
\[
J_{\mathrm{TM}}^{(\varepsilon,\mu)}
=
\mathcal A_L(\varepsilon,\mu)^*
\big(\mathcal A_L(\varepsilon,\mu)\mathcal A_L(\varepsilon,\mu)^*\big)^{-1}
e_{\mathrm{TM},10}.
\]
We define the associated tuning function by
\[
F_{\mathrm{TM}}(\varepsilon,\mu;\rho)
=
\Im\left\{
\frac12\int_V
E[J_{\mathrm{TM}}^{(\varepsilon,\mu)};\varepsilon,\mu]
\cdot
\overline{J_{\mathrm{TM}}^{(\varepsilon,\mu)}}\,d\mathbf r
\right\}.
\]

Because the prescribed datum contains only an electric dipole, the leading quasi-static contribution is controlled by the electric dipole block. The magnetic parameter \(\mu\) enters only through higher-order dynamic corrections.

\begin{proposition}[Single TM dipole leading-order reduction]
\label{prop:tm-dipole-reduction}
Let \(K\subset\Omega_{\rm reg}\cap\mathbb R^2\) be compact and bounded away from \(\varepsilon=0\), \(\mu=0\), \(\varepsilon=-2\), and \(\mu=-2\). There exists a positive scale factor \(S_{\rm e}(\rho)\), independent of \((\varepsilon,\mu)\), and a real-valued function
\[
G_{\rm e}\in C^1(\pi_\varepsilon K)
\]
that depends only on \(\varepsilon\), such that
\[
\frac{F_{\mathrm{TM}}(\varepsilon,\mu;\rho)}{S_{\rm e}(\rho)}
=
G_{\rm e}(\varepsilon)
+
\mathcal O(\rho^2)
\quad\text{in }C^1(K).
\]
The profile \(G_{\rm e}\) is the static electric-dipole profile of the constrained minimum-norm source. It is determined by the ratio of the leading static reactive-power coefficient to the squared leading static exterior-coefficient functional in the TM \(\ell=1\) channel.
\end{proposition}

\begin{proof}
Restrict the truncated forward map and the power quadratic form to the TM \(\ell=1,m=0\) channel. Let \(a_{\rm e}(\varepsilon,\mu;\rho)\) denote the scalar coefficient mapping a normalized electric-dipole source representative to the exterior TM coefficient, and let \(q_{\rm e}(\varepsilon,\mu;\rho)\) denote the corresponding scalar coefficient of \(\Im P\) on the same channel. The small-argument VSWF/Mie expansions give, uniformly with first derivatives on \(K\),
\[
a_{\rm e}(\varepsilon,\mu;\rho)
=
\rho^{p_{\rm e}}
\left(a_{{\rm e},0}(\varepsilon)+\rho^2 a_{{\rm e},2}(\varepsilon,\mu;\rho)\right),
\]
and
\[
q_{\rm e}(\varepsilon,\mu;\rho)
=
\rho^{q_{\rm e}}
\left(q_{{\rm e},0}(\varepsilon)+\rho^2 q_{{\rm e},2}(\varepsilon,\mu;\rho)\right).
\]
The leading coefficient \(a_{{\rm e},0}\) is nonzero on \(K\) because the compact set avoids the static electric dipole pole and the material singularities. For a scalar equality constraint with prescribed exterior coefficient \(\alpha_{10}=1\), the constrained minimum-norm amplitude is proportional to \(\overline{a_{\rm e}}/|a_{\rm e}|^2\). Hence the leading reactive-power contribution is proportional to
\[
\frac{q_{{\rm e},0}(\varepsilon)}{|a_{{\rm e},0}(\varepsilon)|^2}.
\]
Taking
\[
S_{\rm e}(\rho)=\rho^{q_{\rm e}-2p_{\rm e}}
\]
up to a positive normalization constant, and defining \(G_{\rm e}\) by the preceding static ratio, gives the stated expansion. The absence of \(\mu\) from the leading term is the quasi-static electric-dipole decoupling; \(\mu\)-dependence first appears in the \(\rho^2\) correction. Uniformity in \(C^1(K)\) follows from the same small-argument expansions and from the compact exclusion of the singular values.
\end{proof}

\begin{remark}
Proposition~\ref{prop:tm-dipole-reduction} explains why a pure electric-dipole datum is not the best vehicle for displaying a regular two-parameter tuning curve at leading order. The leading term depends only on \(\varepsilon\); any nontrivial two-parameter tuning behavior in the pure TM case is therefore a higher-order or near-resonant effect.
\end{remark}

The dual statement for a pure TE magnetic-dipole datum \(b=e_{\mathrm{TE},10}\) is obtained by electromagnetic duality. There exists a positive scale \(S_{\rm m}(\rho)\) and a real-valued static magnetic profile \(G_{\rm m}\in C^1(\pi_\mu K)\) such that, under the present reactive-power sign convention,
\[
\frac{F_{\mathrm{TE}}(\varepsilon,\mu;\rho)}{S_{\rm m}(\rho)}
=
-\,G_{\rm m}(\mu)
+
\mathcal O(\rho^2)
\quad\text{in }C^1(K).
\]
The minus sign encodes the opposite electric and magnetic reactive signs for the \(e^{-i\omega t}\) convention. Reversing the time convention multiplies the tuning function by an overall sign and does not change the zero set.

\subsection{Mixed electric--magnetic dipole data}
\label{sec:asymptotics:mixed-dipole}

To obtain a genuinely two-parameter leading-order tuning law, we consider mixed electric--magnetic dipole data of the form
\[
b_\lambda=e_{\mathrm{TM},10}+\lambda e_{\mathrm{TE},10},
\qquad \lambda\in\mathbb C,
\]
where \(e_{\mathrm{TE},10}\) denotes the coefficient vector with \(\beta_{10}=1\) and all other exterior coefficients zero. The parameter \(\lambda\) fixes the relative magnetic-dipole content of the prescribed field.

Let \(J_\lambda^{(\varepsilon,\mu)}\) denote the corresponding minimum-energy source,
\[
J_\lambda^{(\varepsilon,\mu)}
=
\mathcal A_L(\varepsilon,\mu)^*
\big(\mathcal A_L(\varepsilon,\mu)\mathcal A_L(\varepsilon,\mu)^*\big)^{-1}
b_\lambda,
\]
and define
\[
F_\lambda(\varepsilon,\mu;\rho)
=
\Im\left\{
\frac12\int_V
E[J_\lambda^{(\varepsilon,\mu)};\varepsilon,\mu]
\cdot
\overline{J_\lambda^{(\varepsilon,\mu)}}\,d\mathbf r
\right\}.
\]
At leading order, the electric and magnetic dipole subspaces are orthogonal under the spherical harmonic decomposition. Cross terms are therefore absent at quasi-static order, and the leading reactive power is the difference between the electric and magnetic static profiles.

\begin{theorem}[Mixed-dipole leading-order tuning law]
\label{thm:mixed-dipole-FLO}
Let \(K\subset\Omega_{\rm reg}\cap\mathbb R^2\) be compact and bounded away from
\(\varepsilon=0\), \(\mu=0\), \(\varepsilon=-2\), and \(\mu=-2\). For each fixed \(\lambda\neq0\), there exists a positive scale factor \(S(\rho,\lambda)\), independent of \((\varepsilon,\mu)\), and a positive dimensionless parameter
\[
\Lambda=\Lambda(\lambda)>0
\]
such that
\[
\frac{F_\lambda(\varepsilon,\mu;\rho)}{S(\rho,\lambda)}
=
F_{\rm LO}(\varepsilon,\mu;\Lambda)
+
\mathcal O(\rho^2)
\quad\text{in }C^1(K),
\]
where the leading-order tuning function is
\[
F_{\rm LO}(\varepsilon,\mu;\Lambda)
\coloneqq
G_{\rm e}(\varepsilon)-\Lambda G_{\rm m}(\mu).
\]
The leading-order tuning curve is therefore
\[
\mathcal M_{\rm LO}^{\Lambda}
=
\left\{
(\varepsilon,\mu)\in K:
G_{\rm e}(\varepsilon)=\Lambda G_{\rm m}(\mu)
\right\}.
\]
\end{theorem}

\begin{proof}
By Proposition~\ref{prop:tm-dipole-reduction}, the electric TM contribution has leading static profile \(G_{\rm e}(\varepsilon)\). By electromagnetic duality, the magnetic TE contribution has leading static profile \(-G_{\rm m}(\mu)\). At quasi-static order the electric and magnetic dipole subspaces are orthogonal: they belong to distinct VSWF families and have opposite parity, so their cross terms in the quadratic form \(P(J;\varepsilon,\mu)\) vanish to leading order. Dynamic electric--magnetic coupling terms enter at order \(\rho^2\).

The relative coefficient multiplying \(G_{\rm m}\) depends on \(|\lambda|^2\) and on the chosen normalizations of the TE and TM coefficient slots. Absorbing these positive constants into \(\Lambda\) and using a common positive scale \(S(\rho,\lambda)\) gives the stated formula. The uniform \(C^1(K)\) remainder follows from the small-argument expansions of the \(\ell=1\) VSWF/Mie coefficients and from the exclusion of the static poles and material singularities on \(K\).
\end{proof}

\begin{remark}[Regular points of the leading-order curve]
The gradient of the reduced leading-order function is
\[
\nabla_{(\varepsilon,\mu)}F_{\rm LO}(\varepsilon,\mu;\Lambda)
=
\left(
G_{\rm e}'(\varepsilon),
-\Lambda G_{\rm m}'(\mu)
\right).
\]
Therefore a point of \(\mathcal M_{\rm LO}^{\Lambda}\) is regular whenever
\[
\big(G_{\rm e}'(\varepsilon),G_{\rm m}'(\mu)\big)\neq(0,0).
\]
Away from such critical points, the leading-order curve is locally \(C^1\). When \(G_{\rm m}'(\mu)\neq0\), it may be written as \(\mu=g(\varepsilon)\) with
\[
g'(\varepsilon)
=
\frac{G_{\rm e}'(\varepsilon)}{\Lambda G_{\rm m}'(\mu)}.
\]
\end{remark}

\subsection{Asymptotic persistence and conditioning}
\label{sec:asymptotics:persistence}

Theorem~\ref{thm:mixed-dipole-FLO} supplies a leading-order function expressed in terms of the static electric and magnetic profiles. We now record the consequence for exact tuning curves at small but nonzero electrical size.

For clarity, define the normalized exact tuning function
\[
\widehat F_{\rho,\lambda}(\varepsilon,\mu)
\coloneqq
\frac{F_\lambda(\varepsilon,\mu;\rho)}{S(\rho,\lambda)} .
\]
Then Theorem~\ref{thm:mixed-dipole-FLO} states that
\[
\widehat F_{\rho,\lambda}
=
F_{\rm LO}(\cdot,\cdot;\Lambda)
+
\mathcal O(\rho^2)
\quad\text{in }C^1(K).
\]

\begin{theorem}[Persistence of mixed-dipole tuning curves]
\label{thm:asymp-persistence}
Let \((\varepsilon_0,\mu_0)\in K\) satisfy
\[
F_{\rm LO}(\varepsilon_0,\mu_0;\Lambda)=0,
\qquad
\nabla_{(\varepsilon,\mu)}F_{\rm LO}(\varepsilon_0,\mu_0;\Lambda)\neq0 .
\]
Then, for all sufficiently small \(\rho\), there exists a point
\((\varepsilon_\rho,\mu_\rho)\in K\) such that
\[
F_\lambda(\varepsilon_\rho,\mu_\rho;\rho)=0
\]
and
\[
\|(\varepsilon_\rho,\mu_\rho)-(\varepsilon_0,\mu_0)\|
\le C\rho^2 .
\]
Moreover, in a neighborhood of \((\varepsilon_0,\mu_0)\), the exact tuning set
\[
\mathcal M_{\rho,\lambda}
=
\{(\varepsilon,\mu):F_\lambda(\varepsilon,\mu;\rho)=0\}
\]
is a \(C^1\) curve and lies within \(O(\rho^2)\) of the leading-order curve
\(\mathcal M_{\rm LO}^{\Lambda}\).
\end{theorem}

\begin{proof}
Since
\[
\nabla F_{\rm LO}(\varepsilon_0,\mu_0;\Lambda)\neq0,
\]
the implicit function theorem applies to the leading-order equation
\[
F_{\rm LO}(\varepsilon,\mu;\Lambda)=0
\]
near \((\varepsilon_0,\mu_0)\). Hence the leading-order zero set is a local \(C^1\) curve.

By Theorem~\ref{thm:mixed-dipole-FLO},
\[
\|\widehat F_{\rho,\lambda}-F_{\rm LO}\|_{C^1(K)}
\le C_1\rho^2 .
\]
Therefore, for sufficiently small \(\rho\), the gradient
\(\nabla\widehat F_{\rho,\lambda}\) remains nonzero in a sufficiently small neighborhood of \((\varepsilon_0,\mu_0)\). Applying the implicit function theorem to \(\widehat F_{\rho,\lambda}\) gives a local \(C^1\) exact tuning curve.

To obtain the \(O(\rho^2)\) distance estimate, take coordinates \((s,n)\) near the leading-order curve, where \(s\) is arclength along the curve and \(n\) is signed normal distance. Since \(\partial_n F_{\rm LO}\) is bounded below in magnitude near the regular point, the scalar equation
\[
\widehat F_{\rho,\lambda}(s,n)=0
\]
has a unique solution \(n_\rho(s)\) for \(s\) near \(s_0\). Moreover,
\[
\widehat F_{\rho,\lambda}(s,0)
=
F_{\rm LO}(s,0)
+
\mathcal O(\rho^2)
=
\mathcal O(\rho^2),
\]
and the derivative with respect to \(n\) is bounded away from zero. The one-dimensional mean-value theorem therefore gives
\[
|n_\rho(s)|\le C_2\rho^2.
\]
This proves the local \(O(\rho^2)\) distance between the exact and leading-order tuning curves. Since multiplication by the positive scale \(S(\rho,\lambda)\) does not change the zero set, the same statement holds for \(F_\lambda\).
\end{proof}

\paragraph{Asymptotic conditioning.}
The normalized leading-order condition number is
\[
\widehat\kappa_{\rm LO}(\varepsilon,\mu;\Lambda)
=
\frac{1}{
\left[
\big(G_{\rm e}'(\varepsilon)\big)^2+
\Lambda^2\big(G_{\rm m}'(\mu)\big)^2
\right]^{1/2}
}.
\]
For the exact normalized problem,
\[
\widehat\kappa_{\rho,\lambda}(\varepsilon,\mu)
=
\frac{1}{\|\nabla_{(\varepsilon,\mu)}\widehat F_{\rho,\lambda}(\varepsilon,\mu)\|}.
\]
The \(C^1\) expansion in Theorem~\ref{thm:mixed-dipole-FLO} implies
\[
\widehat\kappa_{\rho,\lambda}(\varepsilon,\mu)
=
\widehat\kappa_{\rm LO}(\varepsilon,\mu;\Lambda)
+
\mathcal O(\rho^2)
\]
on compact subarcs where \(\|\nabla F_{\rm LO}\|\) is bounded below. Thus the leading-order condition number predicts robust and fragile segments of the exact small-size tuning curve. In terms of the unnormalized tuning function \(F_\lambda\), the condition number differs by a global scale factor, which affects the absolute scale but not the location of fragile segments along a fixed normalized curve.

\section{Numerics: continuation of tuning curves and conditioning maps}
\label{sec:numerics}

The numerical experiments in this section are designed to validate the profile-based asymptotic reduction above. The diagnostics are intentionally organized so that they do not require an assumed closed form for the static profiles \(G_{\rm e}\) and \(G_{\rm m}\). First, a pure TM test verifies that the normalized finite-size response collapses to a one-parameter function of \(\varepsilon\). Second, mixed TM/TE data are used to compute finite-\(\rho\) tuning curves and compare them with the leading-order curve generated by numerically extracted static profiles. Third, conditioning and perturbation experiments validate the robustness interpretation of \(1/\|\nabla F\|\).

\subsection{Continuation procedure}
\label{sec:numerics:continuation}

For a fixed \(\rho\), the finite-\(\rho\) tuning function is evaluated using the dipole-block modal reduction of the Green-kernel formulation. The computation uses the same minimum-norm field-constrained form as the analysis: for each \((\varepsilon,
\mu)\), the source is the Riesz representer of the exterior coefficient constraint, normalized by the corresponding Gram factor. No Lagrange multiplier from the earlier constrained formulation is used as a material parameter.

The exact finite-\(\rho\) mixed tuning curve is computed as the zero contour
\[
F_\lambda(\varepsilon,\mu;\rho)=0
\]
over a compact negative-parameter rectangle away from the excluded singular and resonant values. The leading-order static profiles \(G_{\rm e}\) and \(G_{\rm m}\) are approximated from the smallest computed value of \(\rho\), after confirming that the pure TM and TE responses have negligible spread in the inactive material parameter. This is a numerical representation of the static limits in Proposition~\ref{prop:tm-dipole-reduction}, not an additional fitted closed-form model.

\subsection{Pure TM dipole: one-parameter leading behavior}
\label{sec:numerics:tm-dipole}

\begin{figure}[t]
\centering
\includegraphics[width=0.95\linewidth]{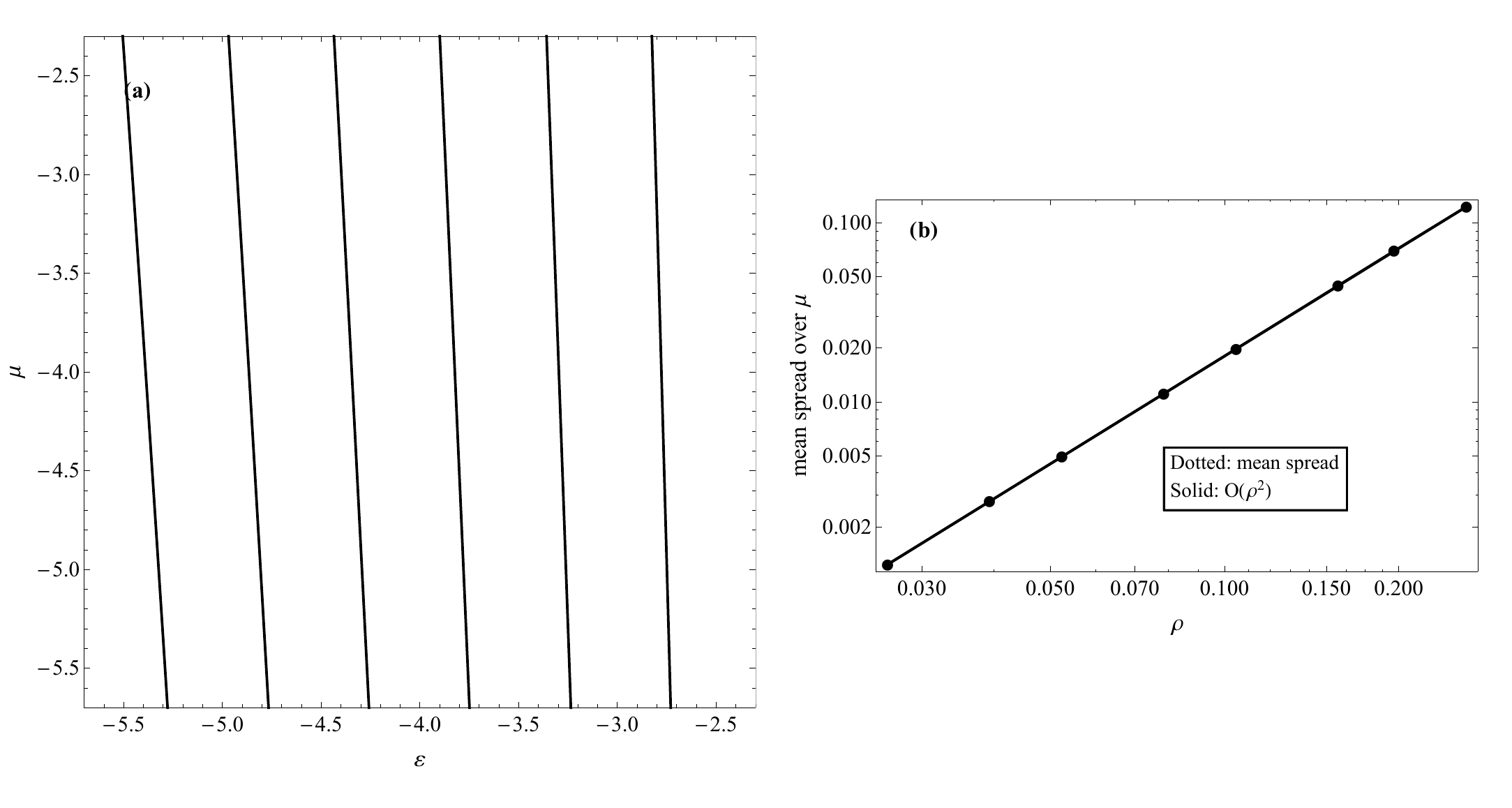}
\caption{Pure TM dipole reduction in the regular negative-parameter window. (a) Normalized finite-\(\rho\) contours of the pure TM modal tuning function for \(b=e_{\mathrm{TM},10}\), showing the dominant one-parameter dependence on \(\varepsilon\). The plotted window is a compact regular subdomain away from \(\varepsilon=0\), \(\mu=0\), and the static resonance values \(\varepsilon=-2\), \(\mu=-2\). (b) Collapse diagnostic for the same pure TM datum. For each \(\rho\), the plotted quantity is the mean spread of the multiplicatively normalized finite-\(\rho\) response over \(\mu\) at fixed \(\varepsilon\). The dashed reference line is proportional to \(\rho^2\). Decay of this spread confirms that the leading static profile is a function of \(\varepsilon\) alone, as predicted by Proposition~\ref{prop:tm-dipole-reduction}.}
\label{fig:tm-dipole-asymptotics}
\end{figure}

The first test uses the single TM electric-dipole datum
\[
b=e_{\mathrm{TM},10}.
\]
According to Proposition~\ref{prop:tm-dipole-reduction}, the normalized small-size tuning function satisfies
\[
\frac{F_{\mathrm{TM}}(\varepsilon,
\mu;\rho)}{S_{\rm e}(\rho)}
=
G_{\rm e}(\varepsilon)+\mathcal O(\rho^2)
\quad\text{in }C^1(K).
\]
Therefore the appropriate numerical test is not level-by-level agreement with a guessed analytic formula, but collapse with respect to the inactive parameter \(\mu\). We measure this by normalizing the finite-\(\rho\) response by a global multiplicative factor and computing the mean spread over \(\mu\) at fixed \(\varepsilon\). Figure~\ref{fig:tm-dipole-asymptotics} reports both diagnostics: panel (a) shows the near-vertical finite-\(\rho\) contours expected from leading dependence on \(\varepsilon\) alone, while panel (b) quantifies the decay of the inactive-\(\mu\) spread against an \(O(\rho^2)\) reference rate.

This experiment is mainly diagnostic. It explains why a pure electric-dipole datum is not the ideal setting for displaying a regular two-parameter tuning curve: the leading-order equation is essentially one-dimensional, and any nontrivial two-parameter tuning behavior in the pure TM case is a higher-order effect.

\subsection{Mixed TM/TE dipole tuning curves}
\label{sec:numerics:mixed-dipole}

The second test uses mixed electric--magnetic dipole data,
\[
b_\lambda=e_{\mathrm{TM},10}+\lambda e_{\mathrm{TE},10}.
\]
For this datum, Theorem~\ref{thm:mixed-dipole-FLO} predicts the normalized expansion
\[
\widehat F_{\rho,\lambda}(\varepsilon,
\mu)
=
F_{\rm LO}(\varepsilon,
\mu;\Lambda)+\mathcal O(\rho^2)
\quad\text{in }C^1(K),
\]
where
\[
F_{\rm LO}(\varepsilon,
\mu;\Lambda)=G_{\rm e}(\varepsilon)-\Lambda G_{\rm m}(\mu).
\]
The leading-order tuning curve is therefore
\[
\mathcal M_{\rm LO}^{\Lambda}
=
\left\{
(\varepsilon,
\mu)\in K:
G_{\rm e}(\varepsilon)=\Lambda G_{\rm m}(\mu)
\right\}.
\]

\begin{figure}[t]
\centering
\includegraphics[width=0.6\linewidth]{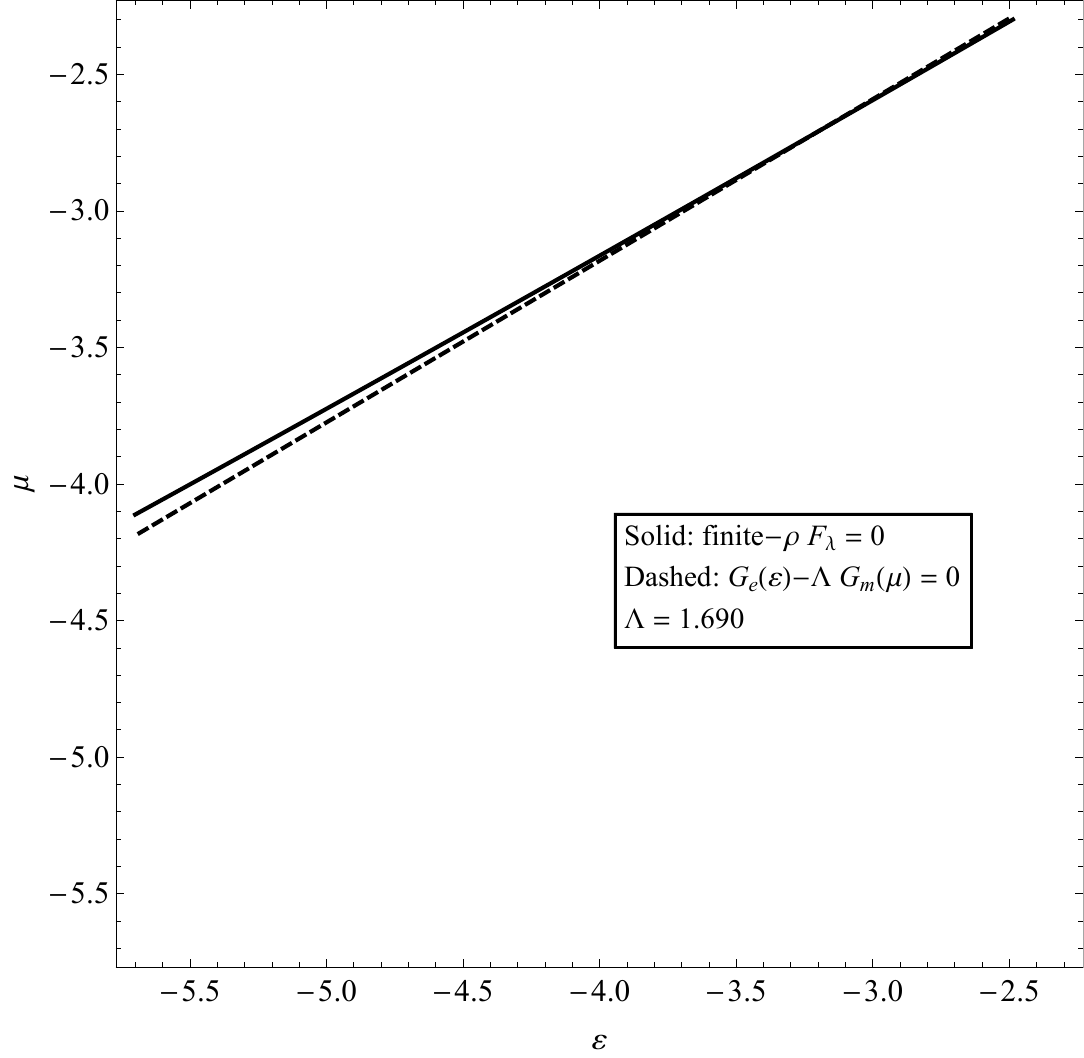}
\caption{Mixed TM/TE dipole tuning curves for \(b_\lambda=e_{\mathrm{TM},10}+\lambda e_{\mathrm{TE},10}\). The solid curve represents the computed finite-\(\rho\) zero set \(F_\lambda(\varepsilon,\mu;\rho)=0\). The dashed curve represents the leading-order prediction \(G_{\rm e}(\varepsilon)=\Lambda G_{\rm m}(\mu)\), with the static profiles \(G_{\rm e}\) and \(G_{\rm m}\) extracted from the smallest-\(\rho\) modal computation. Agreement on the displayed compact regular window illustrates the asymptotic persistence result of Theorem~\ref{thm:asymp-persistence}.}
\label{fig:mixed-dipole-exact-vs-lo}
\end{figure}

Figure~\ref{fig:mixed-dipole-exact-vs-lo} compares this profile-based leading-order curve with the computed finite-\(\rho\) zero contour. The close agreement of the two branches on the displayed compact regular window illustrates the asymptotic persistence mechanism in Theorem~\ref{thm:asymp-persistence}.

\subsection{Conditioning along the tuning manifold}
\label{sec:numerics:conditioning}

\begin{figure}[t]
\centering
\includegraphics[width=0.6\linewidth]{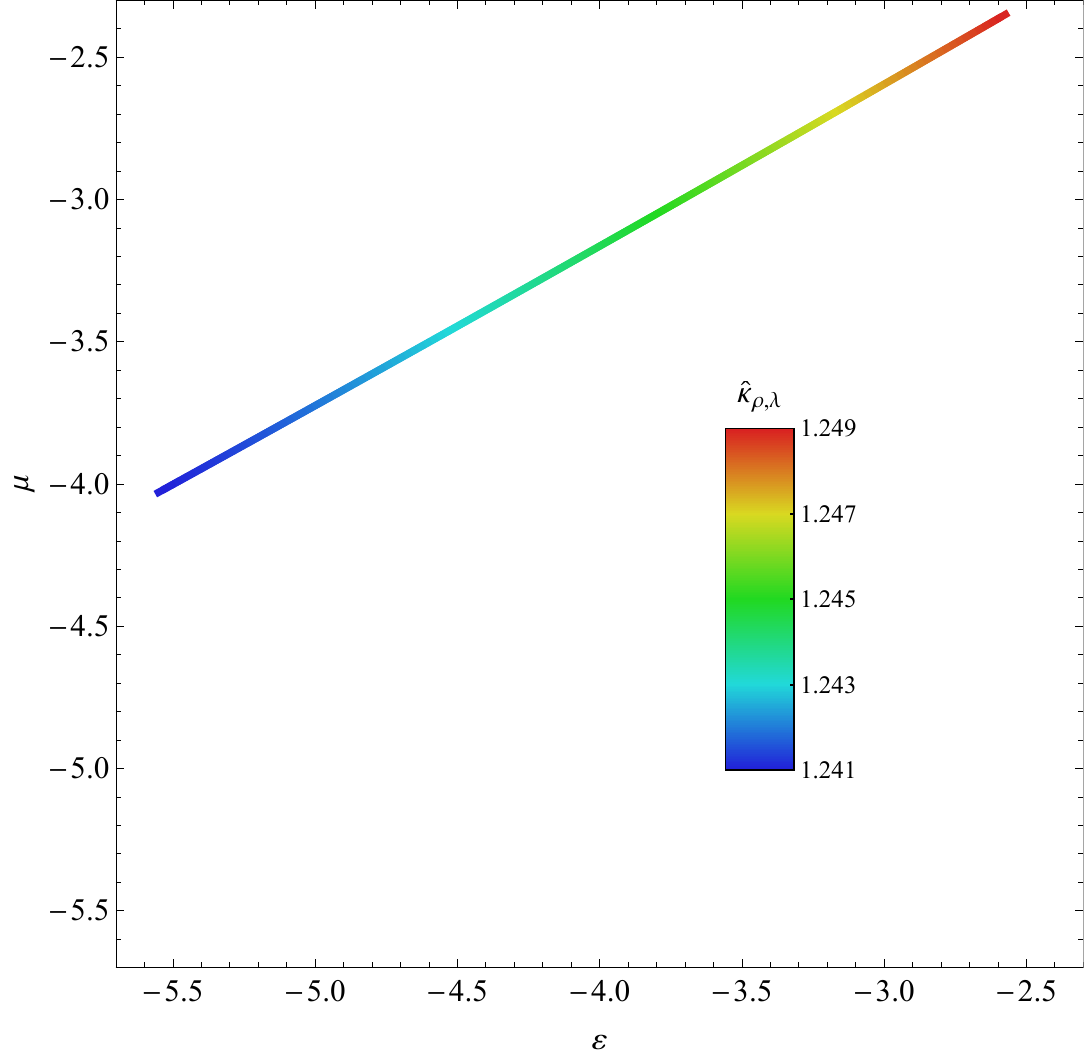}
\caption{Conditioning along the mixed-dipole tuning curve. The color scale represents \(\widehat\kappa_{\rho,
\lambda}(\varepsilon,
\mu)=1/\|\nabla_{(\varepsilon,
\mu)}\widehat F_{\rho,
\lambda}(\varepsilon,
\mu)\|\) evaluated on the computed finite-\(\rho\) tuning curve. Relatively large values identify near-degenerate segments where small material-parameter perturbations produce large normalized reactive-power errors.}
\label{fig:conditioning-curve}
\end{figure}

For the exact mixed-dipole tuning curve, we compute the normalized condition number
\[
\widehat\kappa_{\rho,
\lambda}(\varepsilon,
\mu)
=
\frac{1}{\|\nabla_{(\varepsilon,
\mu)}\widehat F_{\rho,
\lambda}(\varepsilon,
\mu)\|}.
\]
Since the normalization factor multiplying \(F_\lambda\) is positive and independent of \((\varepsilon,
\mu)\), it does not change the location of fragile segments along a fixed curve. Large values of \(\widehat\kappa\) identify regions where the tuning equation is nearly degenerate and where small parameter perturbations can lead to relatively large reactive-power errors. The resulting conditioning distribution along the finite-\(\rho\) branch is shown in Figure~\ref{fig:conditioning-curve}.

\subsection{Perturbation validation}
\label{sec:numerics:perturbation}

\begin{figure}[t]
\centering
\includegraphics[width=0.7\linewidth]{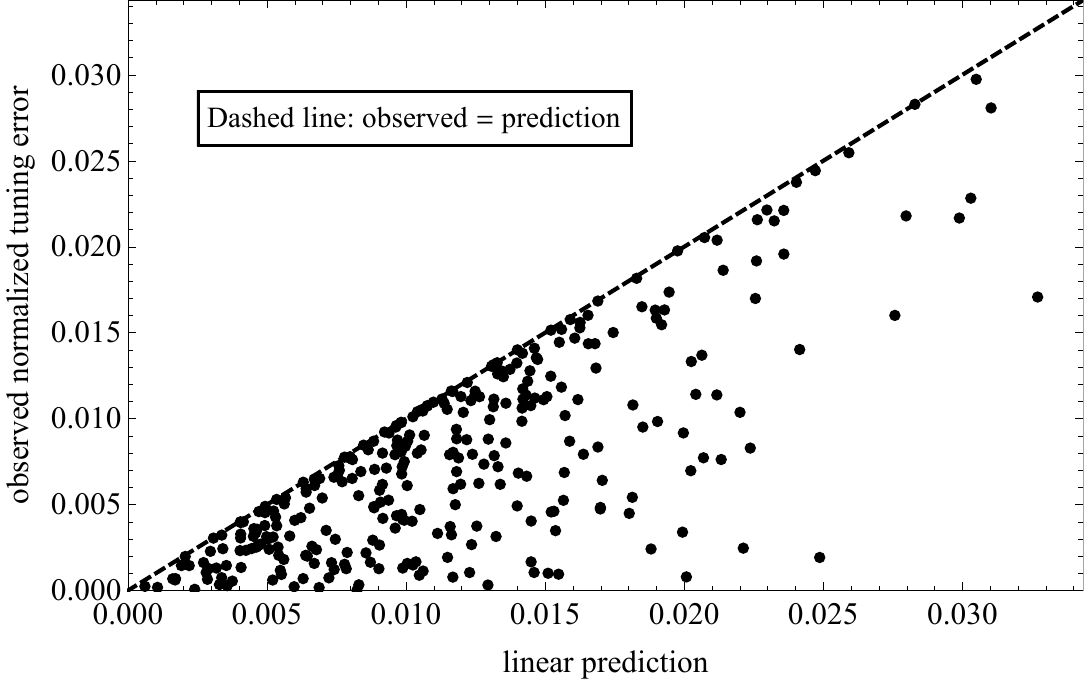}
\caption{First-order robustness validation. For selected regular tuning points \((\varepsilon_0,
\mu_0)\in\mathcal M_{\rho,
\lambda}\), random perturbations \((\delta\varepsilon,
\delta\mu)\) are applied and the observed value of \(|\widehat F_{\rho,
\lambda}(\varepsilon_0+\delta\varepsilon,
\mu_0+\delta\mu)|\) is compared with the linear upper prediction \(\|\nabla_{(\varepsilon,
\mu)}\widehat F_{\rho,
\lambda}(\varepsilon_0,
\mu_0)\|\,\|(\delta\varepsilon,
\delta\mu)\|\). The comparison verifies the local robustness estimate and illustrates how the conditioning metric distinguishes robust and fragile tuned operating points.}
\label{fig:robustness-validation}
\end{figure}

Finally, we test the first-order robustness estimate. Select regular points
\[
(\varepsilon_0,\mu_0)\in\mathcal M_{\rho,
\lambda}
\]
on the computed exact tuning curve and perturb them by
\[
(\varepsilon,
\mu)=(\varepsilon_0,
\mu_0)+(\delta\varepsilon,
\delta\mu).
\]
For sufficiently small perturbations, the Taylor expansion gives
\[
|F_\lambda(\varepsilon_0+\delta\varepsilon,
\mu_0+\delta\mu;\rho)|
\approx
\left|
\nabla_{(\varepsilon,
\mu)}F_\lambda(\varepsilon_0,
\mu_0;\rho)
\cdot
(\delta\varepsilon,
\delta\mu)
\right|
\le
\|\nabla_{(\varepsilon,
\mu)}F_\lambda(\varepsilon_0,
\mu_0;\rho)\|\,
\|(\delta\varepsilon,
\delta\mu)\|.
\]
Thus points with large condition number, equivalently small gradient norm, are expected to show larger tuning errors for a fixed admissible drift. Figure~\ref{fig:robustness-validation} compares the observed normalized residuals with the corresponding first-order upper prediction. The points lying below the dashed reference line are consistent with the local robustness estimate and provide a direct numerical check of the conditioning interpretation used in Figure~\ref{fig:conditioning-curve}.

\subsection{Summary of numerical validation}
\label{sec:numerics:summary}

The experiments summarized in Figures~\ref{fig:tm-dipole-asymptotics}--\ref{fig:robustness-validation} support the analytical picture developed in Section~\ref{sec:asymptotics}. Pure TM dipole data collapse to a one-parameter static profile, with the inactive-parameter spread decreasing as the electrical size is reduced. Mixed TM/TE dipole data produce regular two-parameter tuning curves governed by the balance of the electric and magnetic static profiles. The condition number computed from the exact normalized tuning function identifies robust and fragile segments, and the perturbation experiment verifies the associated first-order robustness estimate. These results provide the numerical basis for using the tuning manifold and its conditioning as a local design and stability diagnostic.

\section{Conclusion}
\label{sec:conclusion}
We characterized the tuning manifold \(\mathcal M=\{\theta:\Im P(\theta)=0\}\) associated with the unique minimum-energy source realizing a prescribed exterior Maxwell field. On the regular parameter set where the truncated forward map has full row rank, we proved differentiability of the optimal-source map and established that \(\mathcal M\) is generically a smooth curve. We then introduced a conditioning perspective that yields quantitative robustness bounds and a practical condition number predicting where tuning is stable or fragile under material-parameter drift. Finally, we formulated a rigorous asymptotic persistence result showing how electrically small tuning curves inherit their structure from leading-order static modal profiles whenever the small-size expansion is uniform in \(C^1\). The numerical experiments demonstrate this profile-based reduction and show how the same framework identifies stable and fragile operating segments along the tuning manifold. Several extensions appear possible, including closed-form characterization of additional static profiles, incorporation of lossy or dispersive material laws, and adaptation of the conditioning framework to settings beyond spherical symmetry.

\section*{Acknowledgment}
The author gratefully acknowledges partial support from Prince Mohammad Bin Fahd University. During preparation of the manuscript, the author used ChatGPT (OpenAI) for drafting assistance, language editing, and code review. The author reviewed, edited, and approved all suggested changes and takes full responsibility for the accuracy and originality of the content.

\appendix
\section{VSWF/VSH conventions and coefficient extraction}
\label{app:vswf}

This appendix fixes the spherical-harmonic, vector-spherical-harmonic, and VSWF normalizations used throughout the paper, and records the coefficient-extraction formulas that define the data map \(\mathcal P_L\).


\subsection{Scalar spherical harmonics and angular operators}
Let \(Y_{\ell m}(\hat{\mathbf r})\) be the (Condon--Shortley) scalar spherical harmonics on \(\mathbb S^2\), orthonormalized as
\[
\int_{\mathbb S^2} Y_{\ell m}(\hat{\mathbf r})\,\overline{Y_{\ell' m'}(\hat{\mathbf r})}\,d\Omega
=\delta_{\ell\ell'}\delta_{mm'}.
\]
Define the angular-momentum operator \(\mathbf L\) acting on sufficiently smooth scalar functions by
\[
\mathbf L \coloneqq -i\,\mathbf r\times\nabla,
\]
so that \(\mathbf L\) is tangential to \(\mathbb S^2\) and depends only on angular derivatives when applied to functions of \(\hat{\mathbf r}\).

\subsection{Vector spherical harmonics}
Define the transverse vector spherical harmonics
\[
\mathbf X_{\ell m}(\hat{\mathbf r}) \coloneqq \frac{1}{\sqrt{\ell(\ell+1)}}\,\mathbf L\,Y_{\ell m}(\hat{\mathbf r}),
\qquad
\mathbf W_{\ell m}(\hat{\mathbf r}) \coloneqq \hat{\mathbf r}\times \mathbf X_{\ell m}(\hat{\mathbf r}),
\]
which satisfy the orthonormality relations
\[
\int_{\mathbb S^2}\mathbf X_{\ell m}\cdot \overline{\mathbf X_{\ell' m'}}\,d\Omega
=
\int_{\mathbb S^2}\mathbf W_{\ell m}\cdot \overline{\mathbf W_{\ell' m'}}\,d\Omega
=\delta_{\ell\ell'}\delta_{mm'},
\qquad
\int_{\mathbb S^2}\mathbf X_{\ell m}\cdot \overline{\mathbf W_{\ell' m'}}\,d\Omega=0.
\]
Both \(\mathbf X_{\ell m}\) and \(\mathbf W_{\ell m}\) are tangential to the sphere.

\subsection{Outgoing vector spherical wave functions (VSWFs)}
Let \(j_\ell\) and \(h_\ell^{(1)}\) denote the spherical Bessel and spherical Hankel functions of the first kind. For \(k>0\), define the outgoing (Hankel-type) VSWFs by
\begin{align}
\mathbf M_{\ell m}^{(3)}(k\mathbf r) &\coloneqq h_\ell^{(1)}(kr)\,\mathbf X_{\ell m}(\hat{\mathbf r}),\label{eq:app:Mdef}\\
\mathbf N_{\ell m}^{(3)}(k\mathbf r) &\coloneqq \frac{1}{k}\nabla\times \mathbf M_{\ell m}^{(3)}(k\mathbf r).\label{eq:app:Ndef}
\end{align}
In a source-free homogeneous region these satisfy
\[
\nabla\times \mathbf M_{\ell m}^{(3)} = k\,\mathbf N_{\ell m}^{(3)},
\qquad
\nabla\times \mathbf N_{\ell m}^{(3)} = k\,\mathbf M_{\ell m}^{(3)},
\qquad
\nabla\cdot \mathbf M_{\ell m}^{(3)}=\nabla\cdot \mathbf N_{\ell m}^{(3)}=0.
\]
For later use, define the standard radial combination
\[
\psi_\ell(z)\coloneqq z\,h_\ell^{(1)}(z),
\qquad
\psi_\ell'(z)=\frac{d}{dz}\big(z\,h_\ell^{(1)}(z)\big).
\]

\subsection{Exterior field expansion and coefficient conventions}
Fix an observation radius \(R>a\). For \(r\ge R\), the radiated field admits the truncated outgoing expansion
\begin{equation}
E(\mathbf r)\big|_{r=R}
=\sum_{\ell=1}^{L}\sum_{m=-\ell}^{\ell}
\Big(
\alpha_{\ell m}\,\mathbf N_{\ell m}^{(3)}(k_0\mathbf r)
+\beta_{\ell m}\,\mathbf M_{\ell m}^{(3)}(k_0\mathbf r)
\Big),
\label{eq:app:Eexpand}
\end{equation}
with associated magnetic field obtained from Maxwell’s equations in vacuum:
\begin{equation}
H(\mathbf r)\big|_{r=R}
=\frac{1}{i\omega\mu_0}\nabla\times E(\mathbf r)\big|_{r=R}
=\frac{1}{\eta_0}\sum_{\ell=1}^{L}\sum_{m=-\ell}^{\ell}
\Big(
\alpha_{\ell m}\,\mathbf M_{\ell m}^{(3)}(k_0\mathbf r)
+\beta_{\ell m}\,\mathbf N_{\ell m}^{(3)}(k_0\mathbf r)
\Big),
\label{eq:app:Hexpand}
\end{equation}
where \(\eta_0=\sqrt{\mu_0/\varepsilon_0}\).

Write tangential traces on \(S_R\) as \(E_t=\hat{\mathbf r}\times(E\times\hat{\mathbf r})\) and \(H_t=\hat{\mathbf r}\times(H\times\hat{\mathbf r})\).
Using \eqref{eq:app:Mdef}--\eqref{eq:app:Ndef}, one has the explicit tangential decompositions at \(r=R\):
\begin{align}
\big(\mathbf M_{\ell m}^{(3)}(k_0\mathbf r)\big)_t\Big|_{r=R} &= h_\ell^{(1)}(k_0R)\,\mathbf X_{\ell m}(\hat{\mathbf r}),\label{eq:app:Mt}\\
\big(\mathbf N_{\ell m}^{(3)}(k_0\mathbf r)\big)_t\Big|_{r=R} &= \frac{\psi_\ell'(k_0R)}{k_0R}\,\mathbf W_{\ell m}(\hat{\mathbf r}).\label{eq:app:Nt}
\end{align}
Therefore, the tangential electric field on \(S_R\) admits the orthogonal VSH decomposition
\begin{equation}
E_t(R\hat{\mathbf r})
=
\sum_{\ell=1}^{L}\sum_{m=-\ell}^{\ell}
\left(
\beta_{\ell m}\,h_\ell^{(1)}(k_0R)\,\mathbf X_{\ell m}(\hat{\mathbf r})
+
\alpha_{\ell m}\,\frac{\psi_\ell'(k_0R)}{k_0R}\,\mathbf W_{\ell m}(\hat{\mathbf r})
\right).
\label{eq:app:EtVSH}
\end{equation}

\subsection{Coefficient extraction and definition of \(\mathcal P_L\)}
Given tangential field traces \(E_t\) on \(S_R\), the orthonormality of \(\mathbf X_{\ell m}\) and \(\mathbf W_{\ell m}\) in \eqref{eq:app:EtVSH} yields the projection formulas
\begin{align}
\beta_{\ell m}
&=
\frac{1}{h_\ell^{(1)}(k_0R)}\int_{\mathbb S^2} E_t(R\hat{\mathbf r})\cdot \overline{\mathbf X_{\ell m}(\hat{\mathbf r})}\,d\Omega,\label{eq:app:beta}\\
\alpha_{\ell m}
&=
\frac{k_0R}{\psi_\ell'(k_0R)}\int_{\mathbb S^2} E_t(R\hat{\mathbf r})\cdot \overline{\mathbf W_{\ell m}(\hat{\mathbf r})}\,d\Omega.\label{eq:app:alpha}
\end{align}
Equations \eqref{eq:app:beta}--\eqref{eq:app:alpha} define the coefficient-extraction operator
\[
\mathcal P_L:\ E_t|_{S_R}\ \mapsto\ b=(\{\alpha_{\ell m}\},\{\beta_{\ell m}\})\in\mathbb C^{N_L},
\qquad N_L=2\sum_{\ell=1}^{L}(2\ell+1)=2L(L+2),
\]
with the ordering of entries fixed once and used consistently throughout.

\begin{remark}[Alternative extraction using \((E_t,H_t)\)]
In numerics, one may equivalently compute \(\alpha_{\ell m}\) from \(H_t\) via \eqref{eq:app:Hexpand} and the analog of \eqref{eq:app:beta}--\eqref{eq:app:alpha}. Using both \(E_t\) and \(H_t\) can improve conditioning when \(|h_\ell^{(1)}(k_0R)|\) or \(|\psi'_\ell(k_0R)|\) is small for certain \(\ell\) at the chosen \(R\).
\end{remark}

\section{Adjoint operators and synthesis forms for \(\mathcal P_L\) and \(\mathcal A_L(\theta)\)}
\label{app:operators}

This appendix records (i) the adjoint \(\mathcal P_L^*\) of the coefficient-extraction map \(\mathcal P_L\) (equivalently, a synthesis map from coefficients back to tangential fields on \(S_R\)), and (ii) the resulting factorization of the adjoint \(\mathcal A_L(\theta)^*\) and Gram matrix \(G_L(\theta)=\mathcal A_L(\theta)\mathcal A_L(\theta)^*\).

\subsection{Inner products}
We use the standard complex Hilbert inner products
\[
\langle u,v\rangle_{\mathbb C^{N_L}} \coloneqq \sum_{j=1}^{N_L} u_j\,\overline{v_j},
\qquad
\langle f,g\rangle_{S_R} \coloneqq \int_{\mathbb S^2} f(R\hat{\mathbf r})\cdot \overline{g(R\hat{\mathbf r})}\,d\Omega,
\]
where \(f,g\in L^2_{\mathrm{tan}}(S_R)^2\) are tangential vector fields on \(S_R\) and \(d\Omega\) is the solid-angle measure.
The current space is \(\mathcal H=L^2(V)^3\) with \(\langle J_1,J_2\rangle_{\mathcal H}=\int_V J_1\cdot \overline{J_2}\,d\mathbf r\).

\subsection{Synthesis operator \(\mathcal P_L^*\)}
Write the truncated coefficient vector as
\[
y=\Big(\{u_{\ell m}\}_{1\le \ell\le L,|m|\le \ell},\ \{v_{\ell m}\}_{1\le \ell\le L,|m|\le \ell}\Big)\in\mathbb C^{N_L},
\]
where \(u_{\ell m}\) corresponds to the \(\alpha_{\ell m}\) (TM/\(\mathbf N\)-family) slot and \(v_{\ell m}\) to the \(\beta_{\ell m}\) (TE/\(\mathbf M\)-family) slot.
With the extraction formulas \eqref{eq:app:beta}--\eqref{eq:app:alpha} and the orthonormality of \(\mathbf X_{\ell m}\), \(\mathbf W_{\ell m}\), the adjoint \(\mathcal P_L^*: \mathbb C^{N_L}\to L^2_{\mathrm{tan}}(S_R)^2\) is characterized by
\[
\langle \mathcal P_L E_t,\ y\rangle_{\mathbb C^{N_L}}
=
\langle E_t,\ \mathcal P_L^* y\rangle_{S_R}
\qquad \text{for all }E_t\in L^2_{\mathrm{tan}}(S_R)^2.
\]
A direct substitution of \eqref{eq:app:beta}--\eqref{eq:app:alpha} gives the explicit synthesis form
\begin{equation}
\big(\mathcal P_L^* y\big)(R\hat{\mathbf r})
=
\sum_{\ell=1}^{L}\sum_{m=-\ell}^{\ell}
\left(
\frac{v_{\ell m}}{h_\ell^{(1)}(k_0R)}\,\mathbf X_{\ell m}(\hat{\mathbf r})
+
\frac{k_0R}{\psi_\ell'(k_0R)}\,u_{\ell m}\,\mathbf W_{\ell m}(\hat{\mathbf r})
\right).
\label{eq:PLstar}
\end{equation}
Thus \(\mathcal P_L^*\) maps coefficient weights to the corresponding tangential VSH superposition on \(S_R\), with the same radial normalization factors that appear in the extraction formulas.

\subsection{Adjoint factorization for \(\mathcal A_L(\theta)\) and the Gram matrix}
Recall that the truncated data operator is the composition
\[
\mathcal A_L(\theta)=\mathcal P_L\circ \mathcal S(\theta):\ \mathcal H\to\mathbb C^{N_L},
\]
where \(\mathcal S(\theta):\mathcal H\to L^2_{\mathrm{tan}}(S_R)^2\) maps a source current \(J\) to the tangential electric trace \(E_t|_{S_R}\) of the radiated field solving Maxwell’s equations in the two-region medium.
With the inner products above, the adjoint satisfies the standard composition rule
\begin{equation}
\mathcal A_L(\theta)^*=\mathcal S(\theta)^*\circ \mathcal P_L^*:\ \mathbb C^{N_L}\to \mathcal H,
\label{eq:ALstar}
\end{equation}
where \(\mathcal S(\theta)^*:L^2_{\mathrm{tan}}(S_R)^2\to \mathcal H\) is the Hilbert adjoint of the Maxwell trace map.

Consequently, the Gram matrix/operator governing existence, uniqueness, and conditioning of the minimum-energy solution is
\begin{equation}
G_L(\theta)\coloneqq \mathcal A_L(\theta)\mathcal A_L(\theta)^*
=\mathcal P_L\,\mathcal S(\theta)\mathcal S(\theta)^*\,\mathcal P_L^*
:\ \mathbb C^{N_L}\to \mathbb C^{N_L}.
\label{eq:GramFactor}
\end{equation}
In particular, the explicit formula
\[
J^{(\theta)}=\mathcal A_L(\theta)^*\,G_L(\theta)^{-1}\,b
\]
used in the main text is well-defined precisely when \(G_L(\theta)\) is invertible.

\subsection{Practical computation of adjoints}
In numerical implementations, one typically constructs a finite-dimensional discretization of the source space (basis \(\{\varphi_j\}\subset\mathcal H\)) and assembles the complex matrix
\[
A_L(\theta)\in\mathbb C^{N_L\times n}
\quad\text{with columns}\quad
A_L(\theta)[:,j]=\mathcal A_L(\theta)\varphi_j,
\]
either via analytic coupling coefficients (in spherical settings) or via a Maxwell solver producing \(E_t|_{S_R}\) followed by \(\mathcal P_L\).
Then the discrete adjoint is simply the Hermitian transpose \(A_L(\theta)^{H}\), and the discrete Gram matrix is \(G_L(\theta)=A_L(\theta)A_L(\theta)^H\).
Equation \eqref{eq:PLstar} remains useful as an internal consistency check: \(\mathcal P_L^*\) provides a direct way to synthesize tangential test fields from coefficient vectors, which can be backpropagated by an adjoint Maxwell solve if one chooses to realize \(\mathcal S(\theta)^*\) at the PDE level.

\subsection{PDE-level adjoint for the Maxwell trace map $\mathcal S(\theta)$}
\label{app:pdeadjoint}

This subsection records an adjoint-state characterization of \(\mathcal S(\theta)^*\), included for completeness and for extensions beyond the spherical/modal setting.
We keep the time convention \(e^{-i\omega t}\).

\paragraph{State problem.}
Let \(J\in \mathcal H=L^2(V)^3\) be supported in \(V\), and let \((E,H)\) solve
\begin{align}
\nabla\times E &= i\omega \mu(\mathbf r)\,H,\label{eq:adj_state1}\\
\nabla\times H &= -\,i\omega \varepsilon(\mathbf r)\,E + J,\label{eq:adj_state2}
\end{align}
in \(\mathbb R^3\) with outgoing radiation condition, where \(\varepsilon(\mathbf r),\mu(\mathbf r)\) are the piecewise-constant medium fields associated with \(\theta=(\varepsilon,\mu)\) as in Section~\ref{sec:model:maxwell}.
Define the (tangential) trace operator \(\gamma_t\) on \(S_R\) by \(\gamma_t E \coloneqq E_t=\hat{\mathbf r}\times(E\times\hat{\mathbf r})\).
Then the Maxwell trace map used in the main text is \(\mathcal S(\theta)J=\gamma_t E\in L^2_{\mathrm{tan}}(S_R)^2\).

\paragraph{Adjoint problem.}
Let \(g\in L^2_{\mathrm{tan}}(S_R)^2\) be a tangential test function (a ``data residual'' in inverse-problems language).
Define an adjoint field pair \((E^\dagger,H^\dagger)\) as the unique outgoing solution of the homogeneous Maxwell system
\begin{align}
\nabla\times E^\dagger &= -\,i\omega \overline{\mu(\mathbf r)}\,H^\dagger,\label{eq:adj_pde1}\\
\nabla\times H^\dagger &= \ \ i\omega \overline{\varepsilon(\mathbf r)}\,E^\dagger,\label{eq:adj_pde2}
\end{align}
in \(\mathbb R^3\setminus S_R\), with the interface condition on \(S_R\)
\begin{equation}
\hat{\mathbf r}\times H^\dagger\Big|_{S_R} = g,
\label{eq:adj_bc}
\end{equation}
and continuity of the tangential fields across \(S_R\) in the distributional sense. (Equivalently, \eqref{eq:adj_bc} can be realized by an infinitesimal electric surface-current excitation on \(S_R\).)

\begin{proposition}[Adjoint-state identity for \(\mathcal S(\theta)^*\)]
\label{prop:adjointstate}
Assume the state and adjoint problems above are well-posed for the given \(\theta\).
Then the Hilbert adjoint \(\mathcal S(\theta)^*:L^2_{\mathrm{tan}}(S_R)^2\to \mathcal H\) is given by
\begin{equation}
\big(\mathcal S(\theta)^* g\big)(\mathbf r) = -\, \overline{E^\dagger(\mathbf r)}\Big|_{V},
\label{eq:Sstar}
\end{equation}
in the sense that for all \(J\in\mathcal H\) one has the duality relation
\begin{equation}
\langle \mathcal S(\theta)J,\ g\rangle_{S_R}
=
\langle J,\ \mathcal S(\theta)^*g\rangle_{\mathcal H}.
\label{eq:duality}
\end{equation}
\end{proposition}

\begin{proof}[Proof sketch]
Multiply \eqref{eq:adj_state2} by \(\overline{E^\dagger}\), \eqref{eq:adj_pde2} by \(E\), subtract, and use vector Green identities over the ball of radius \(R\) (and its exterior complement), together with \eqref{eq:adj_state1}--\eqref{eq:adj_pde1} to eliminate volume terms in \(\varepsilon,\mu\).
The remaining boundary term on \(S_R\) reduces to \(\int_{S_R} (\gamma_t E)\cdot \overline{g}\,dS\) upon imposing \(\hat{\mathbf r}\times H^\dagger=g\).
The source term yields \(\int_{V} J\cdot \overline{E^\dagger}\,d\mathbf r\), which gives \eqref{eq:Sstar} under the inner-product conventions of Appendix~\ref{app:operators}.
\end{proof}

\begin{remark}[Adjoint for the composed map \(\mathcal A_L(\theta)\)]
Since \(\mathcal A_L(\theta)=\mathcal P_L\circ \mathcal S(\theta)\), the adjoint factorization
\[
\mathcal A_L(\theta)^*=\mathcal S(\theta)^*\circ \mathcal P_L^*
\]
in \eqref{eq:ALstar} becomes constructive: given \(y\in\mathbb C^{N_L}\), first synthesize a tangential field \(g=\mathcal P_L^*y\) on \(S_R\) using \eqref{eq:PLstar}, then solve the adjoint Maxwell problem \eqref{eq:adj_pde1}--\eqref{eq:adj_bc} to obtain \(E^\dagger\), and finally restrict \(-\overline{E^\dagger}\) to \(V\) to recover \(\mathcal A_L(\theta)^*y\).
In the purely modal spherical setting, this procedure reproduces the matrix adjoint \(A_L(\theta)^H\).
\end{remark}

\begin{remark}[On complex parameters and active models]
The complex conjugation in \eqref{eq:adj_pde1}--\eqref{eq:adj_pde2} is a standard consequence of taking adjoints in complex Hilbert spaces.
For passive lossy media, \(\Im \varepsilon\ge 0\), \(\Im \mu\ge 0\); effective active models correspond to allowing negative imaginary parts in some frequency bands, in which case well-posedness must be interpreted within the adopted (stability/causality-consistent) modeling framework.
\end{remark}

\bibliographystyle{unsrt}

\end{document}